\documentclass[preprint,epsfig]{revtex4}

\usepackage{graphicx}
\usepackage{amssymb}
\usepackage{amsmath}
\usepackage{color}
\usepackage{bm}
\usepackage{physics}
\usepackage{changes}
\usepackage{gensymb}

\newcommand{\figsizeone}{0.9}

\begin{document}

\draft
\title{Transformation cavities with a narrow refractive index profile}
\author{Jung-Wan Ryu}
\email{jungwanryu@gmail.com}
\affiliation{Center for Theoretical Physics of Complex Systems, Institute for Basic Science, Daejeon 34051, Republic of Korea}
\author{Jinhang Cho}
\affiliation{Digital Technology Research Center, Kyungpook National University, Daegu 41566, Republic of Korea}

\begin{abstract}
Recently, gradient index cavities, or so-called transformation cavities, designed by conformal transformation optics have been studied to support resonant modes with both high Q-factors and emission directionality. We propose a new design method for transformation cavities to realize a narrower width of the refractive index profile, a great advantage in experimental implementations, without losing the benefits of conformal mapping. We study the resonant modes with both high Q-factor and directional emission in the newly designed transformation cavities, where the refractive index profile is $50 \%$ narrower than in previously proposed transformation cavities. By varying a system parameter with a fixed maximal value of the refractive index profile inside the cavity, the width of the refractive index profile narrows, the Q-factors become higher, and the near and far field patterns maintain their properties, namely conformal whispering gallery modes and bidirectional emission, respectively.
\end{abstract}

\maketitle
\section{Introduction}

Transformation optics (TO) based on the concept that Maxwell's equations can be written in a form-invariant manner under coordinate transformation provides a new way to design optical devices that exhibit extraordinary phenomena \cite{Leo06,Pen06}. A curved spatial geometry of the propagating electromagnetic waves in real space can be realized by modified permittivity and permeability tensors following coordinate transformation. TO has developed many unprecedented applications such as invisibility cloaks \cite{Pen06,Sch06,Cai07,Lan13}, wave concentrators \cite{Jia08,Sad15}, optical black holes \cite{Gen09,Che10,Deh15}, waveguide devices \cite{Rob08,Tic10}, and negative refractive index materials \cite{Ves68,She01,Zha16}.

The transformation cavity (TC), which is a deformed gradient index microcavity designed employing TO, has been proposed to obtain directional light emission while simultaneously maintaining the nature of high-Q whispering gallery modes (WGMs) \cite{Kim16}. The boundary shapes and corresponding refractive index profiles of TCs have been designed utilizing conformal TO \cite{Xu14}. TCs have attracted considerable attention not only in the field of resonator optics, combining optical microcavities with TO, but also in applications requiring high-Q modes with directional light emission. In a limaçon-shaped TC, for example, the optimal set of system parameters within an attainable range of refractive indices for the resonant modes having both high Q-factor and strong bidirectional emission has been obtained \cite{Ryu19}. Moreover, a quasi-conformal mapping method has extended the applicability of TCs to arbitrary shapes \cite{Par19}, and center-shift circular TCs have been introduced as the simplest boundary shapes \cite{Cho20}.

Recently, metamaterials have been intensively studied in various fields due to their extraordinary characteristics \cite{Ves68,Pen99,Smi00,Eng06,Zhe12,Jah16}. The unconventional material parameter values of metamaterials arise from the subwavelength nature of their metallic or dielectric structure and the local resonances of such building blocks. Composites with subwavelength-scale artificial structures can be regarded as homogeneous effective media, as derived from their minimal wave scattering. The local resonances of the local structures lead to extraordinary effective medium parameter values that are rarely or never found in nature \cite{Smi04}. In natural materials, it is extremely difficult to spatially control the refractive index value, and even when possible, the controllable range of refractive index profiles is highly limited. In this light, if we use subwavelength structures such as holes and posts to produce homogenized effective media in the metamaterial regime, the spatial profile of the refractive index can be easily controlled within a wide range. Metamaterials with air holes or dielectric posts related to applications of TO theory such as optical cloaking have been widely studied both experimentally and theoretically \cite{Val09,Gab09,Vas10,Gao12}; as a result, metamaterials have emerged as a promising candidate for the implementation of TCs as well as for other applications of TO theory. In one related work, for instance, the associated conformal WGM (cWGM) of an implemented TC was experimentally observed at microwave frequencies using alumina posts as the subwavelength structure \cite{Kim16}.

Despite such progress, it remains difficult to implement a TC laser at optical frequencies because of its wide range of refractive index profiles, which presents an issue because a deformed cavity with many subwavelength holes is a promising candidate to realize TC lasers. Here, the wider the range of refractive index profiles, the larger the number of holes, the latter of which leads to two inevitable problems: a reduction in the solidity of the cavity, and a decrease in the proportion of gain material. That is, with a large number of holes, the cavity can be easily broken and lasing requiring sufficient gain is not easy to achieve. Additionally, there is an unavoidable limited range of refractive index profiles. The maximal values of the refractive index cannot be larger than the refractive index of the material. As a result, the first process in designing a TC should be the decision of the maximal values of the TC refractive index profile reflecting the material properties. Next, the refractive index profiles, corresponding to boundary shapes, with fixed maximal values should be obtained as narrow as possible, while maintaining the superior optical properties such as high Q-factor and directional emission. In this work, we propose a new design method for a TC that achieves both high Q-factor and emission directionality under the condition of a fixed maximal refractive index value with a narrow refractive index profile. We study the optical properties, namely the Q-factors and the near and far field patterns, of the resonant modes in the newly designed TC.

\section{Systems}
If we consider an infinite cylindrical dielectric cavity with translational symmetry along the $z$-axis, the Maxwell equations are reduced to two-dimensional (2D) scalar wave equations. In this case, one can use an effective 2D dielectric cavity model, where the optical modes are described by resonances or quasibound modes that are obtained by solving the following scalar wave equation,
\begin{equation}
[\nabla^2 + n^2 (\mathbf{r}) k^2] \psi (\mathbf{r}) = 0,
\end{equation}
where $n(\mathbf{r})$ is the refractive index function and $\mathbf{r}=(x, y)$. The resonant modes should satisfy the outgoing-wave boundary condition.
In a conventional deformed cavity with a homogeneous refractive index profile, $n(\mathbf{r})$ is $n_0$ inside the cavity and $1$ outside the cavity. In the case of transverse magnetic (TM) polarization, the wave function $\psi(\mathbf{r})$ corresponds to $E_z$, the $z$-component of the electric field \cite{Jac62}. Both the wave function $\psi(\mathbf{r})$ and its normal derivative $\partial_\nu \psi$ are continuous across the cavity boundary in the case of TM polarization. Here, we focus on TM polarization modes. The real part of the complex wave number $k$ is equal to $\omega / c$, where $\omega$ is the frequency of the resonant mode and $c$ is the speed of light. The imaginary part of $k$ is equal to $- 1/(2 c \tau)$, where $\tau$ is the lifetime of the mode. The quality factor $Q$ of the mode is defined as $Q = 2 \pi \tau / T = - \mathrm{Re}(k) / 2 \mathrm{Im}(k)$.

\begin{figure}[tb]
\centering
\fbox{\includegraphics[width=\figsizeone\linewidth]{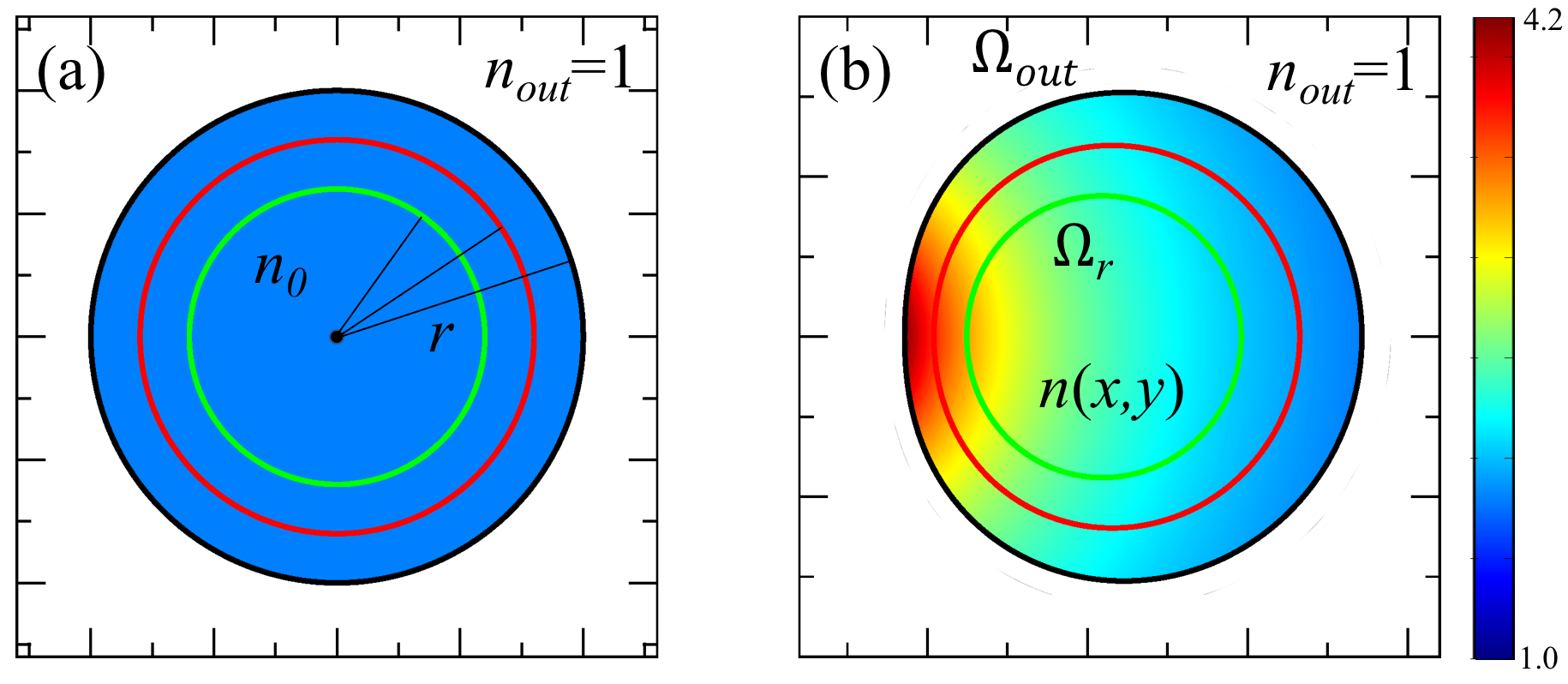}}
\caption{Boundary shapes with $r=1.0$ (black), $r=0.8$ (red), and $r=0.6$ (green) and refractive indices (scaled color) for (a) circular cavities in an original virtual space and (b) corresponding lima\c{c}on-shaped transformation cavities in a physical space when $\epsilon=0.2$. The regions inside and outside the cavities are $\Omega_r$ and $\Omega_{out}$, respectively, for the cavities with different boundary shapes depending on $r$. The refractive indices of the lima\c{c}on-shaped transformation cavities are $n (x,y)$. The white background represents a refractive index of $n_{out}=1$ in the air when $r=1.0$. }
\label{fig1}
\end{figure}

In an original virtual (OV) space or reference space [Fig.~\ref{fig1} (a)], a cavity has a homogeneous refractive index and the resonant modes are the solution of Maxwell's equations in linear isotropic dielectric media. Maxwell's equations maintain their invariant forms under conformal transformation, by which the OV space is transformed into a physical space, as in Fig.~\ref{fig1} (b). In a physical space, the cavity boundary is distorted and the inhomogeneous profile of the refractive index over the whole space can be obtained by the ratio between the local length scales in both spaces. To create a TC in physical space as in Fig.~\ref{fig1} (b), the refractive index outside the cavity is set to $1$; that is, conformal mapping is applied only to the inside of the cavity, rather than to both the inside and outside of the cavity as in conventional TO. This unique setting produces novel properties of the resonant modes in the TCs.

\begin{figure}[tb]
\centering
\fbox{\includegraphics[width=\figsizeone\linewidth]{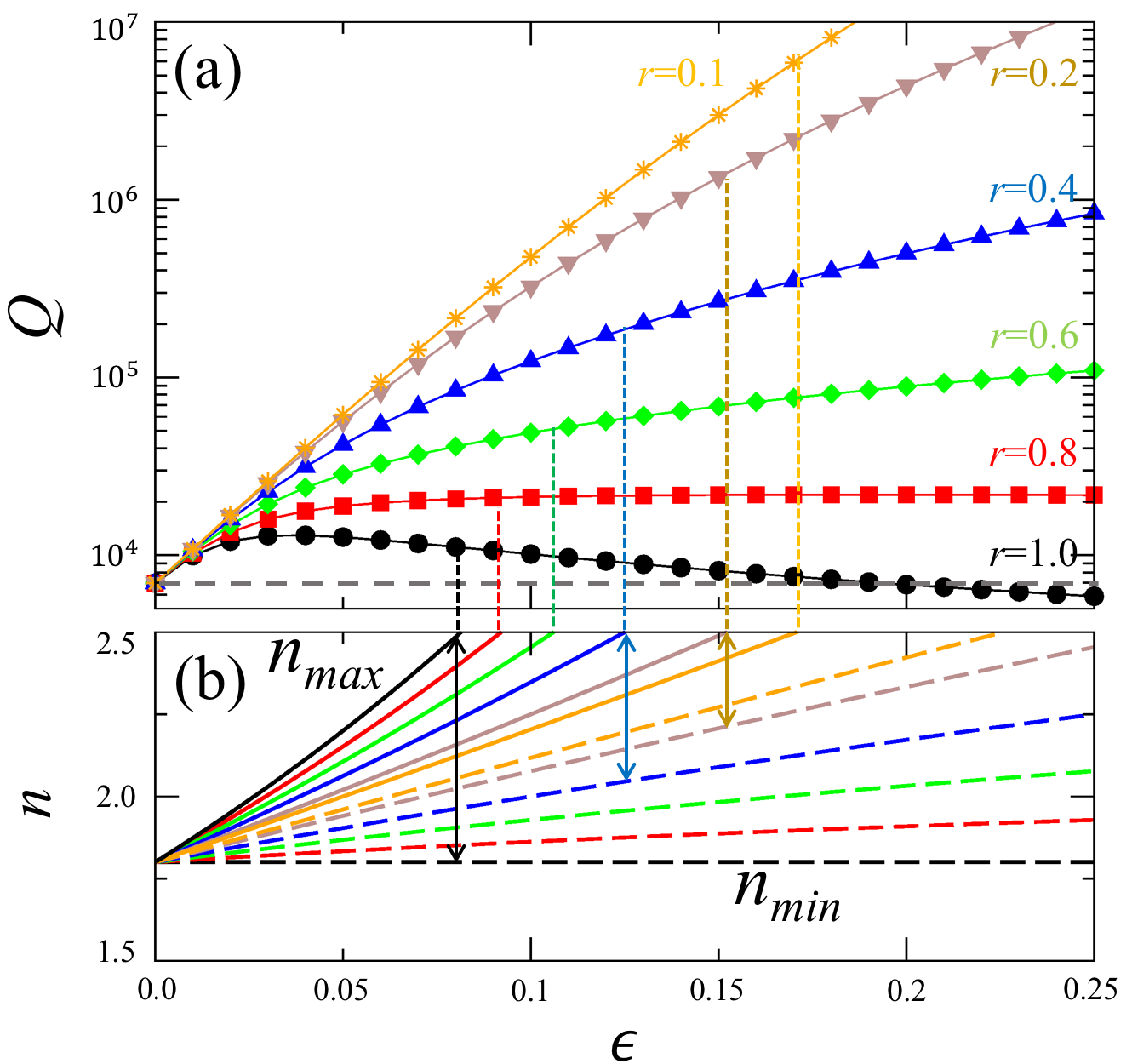}}
\caption{(a) Q-factors of the resonant modes in the lima\c{c}on-shaped transformation cavities as a function of $\epsilon$ for $r=1.0$ (black), $r=0.8$ (red), $r=0.6$ (green), $r=0.4$ (blue), $r=0.2$ (brown), and $r=0.1$ (orange). The gray dashed line indicates the Q-factor of the corresponding WGM in the circular cavity with $n=1.8$ in the OV space. (b) Maximal (solid lines) and minimal (long dashed lines) values of the refractive indices, $n_{max}$ and $n_{min}$, as a function of $\epsilon$ for different $r$. The vertical dashed lines indicate the value of $\epsilon$ when $n_{max} = 2.5$ with different $r$. The black, blue, and brown double arrows represent the widths of the refractive index profiles inside the TCs with $n_{max} = 2.5$ for $r=1.0$, $r=0.4$, and $r=0.2$, respectively.
}
\label{fig2}
\end{figure}

As above, conformal mapping is applied only to the inside of the cavity, denoted by $\Omega_r$ in Fig.~\ref{fig1} (b) where $r$ is the radius of a circular cavity in OV space. As an example, we consider a TC with boundaries characterized by a lima\c{c}on shape, which is one of the widely studied shapes in the fields of quantum billiards \cite{Rob83}, deformed microcavities \cite{Wie08}, and transformation cavities \cite{Kim16}. The corresponding conformal mapping from unit circle to lima\c{c}on is given by
\begin{equation}
\zeta = \beta (\eta + \epsilon \eta^2),
\label{conformal}
\end{equation}
where $\eta = u + i v$ and $\zeta = x + i y$ are complex variables denoting positions in the OV space in Fig.~\ref{fig1} (a) and the physical space in Fig.~\ref{fig1} (b), respectively, $\epsilon$ is a deformation parameter, and $\beta$ is a positive size-scaling parameter. The boundary shapes are given by
\begin{eqnarray}
\label{boundary}
x &=& \beta_{max}(r \cos(\phi) + \epsilon r^2 \cos(2 \phi)),\\\nonumber
y &=& \beta_{max}(r \sin(\phi) + \epsilon r^2 \sin(2 \phi)),
\end{eqnarray}
where size scale factor $r$ is equivalent to the radii of three circles of a concentric circular cavity in an OV space [Fig.~\ref{fig1} (a)], respectively, and $\beta_{max} = 1/(1+2\epsilon)$. The maximal value $\beta_{max}$ of $\beta$, that supports the cWGMs in a limaçon-shaped TC with $r=1.0$, can be obtained from the condition $|d{\zeta} / d{\eta} |^{-1}  \geq 1$ necessary for total internal reflection in the TCs with an outside refractive index of $n_{out} = 1$. The refractive index profiles are given by
\begin{eqnarray}
\label{indices}
n(x,y) =
\begin{cases}
 n_0 \left| \frac{d \zeta}{d \eta} \right|^{-1} = \frac{n_0}{\beta_{max}\left|\sqrt{1+4 \epsilon \zeta / \beta_{max}}\right|} ~~~~& \mathrm{for}~ (x, y) \in \Omega_{r}, \\
 1 ~~~~& \mathrm{for}~ (x, y) \in \Omega_{out},
\end{cases}
\end{eqnarray}
where $n_0$ is the refractive index of the disk cavity in OV space. Throughout this work, $n_0 = 1.8$. The difference between our new design and the original design for a TC ($r=1.0$) is that the boundary between $\Omega_r$ and $\Omega_{out}$ changes depending on the parameter $r$. The important point here is that there is no change in the refractive index profiles inside the cavities even though the boundary shapes are changed. It is noted that this new cavity boundary is also lima\c{c}on shaped with rescaled parameters, i.e., $\epsilon = \epsilon r$ and $\beta = \beta_{max} r$, and also that the newly designed TC preserves the characteristics of conformal mapping. In this cavity, the maximal and minimal values of the refractive indices can be obtained from Eq.~(\ref{boundary}) and Eq.~(\ref{indices}) as follows,
\begin{equation}
n(x,y) = \frac{n_{0} (1+2 \epsilon)}{\left| 1\mp 2 \epsilon r \right|},
\label{maxmin_index}
\end{equation}
where minus and plus signs indicate the maximal and minimal values at
\begin{equation}
(x,y) = \left(\frac{r (r \epsilon \mp 1)}{1+ 2 \epsilon},0\right),
\label{maxmin_position}
\end{equation}
respectively.

\section{Resonant modes}

\begin{figure}[tb]
\centering
\fbox{\includegraphics[width=\figsizeone\linewidth]{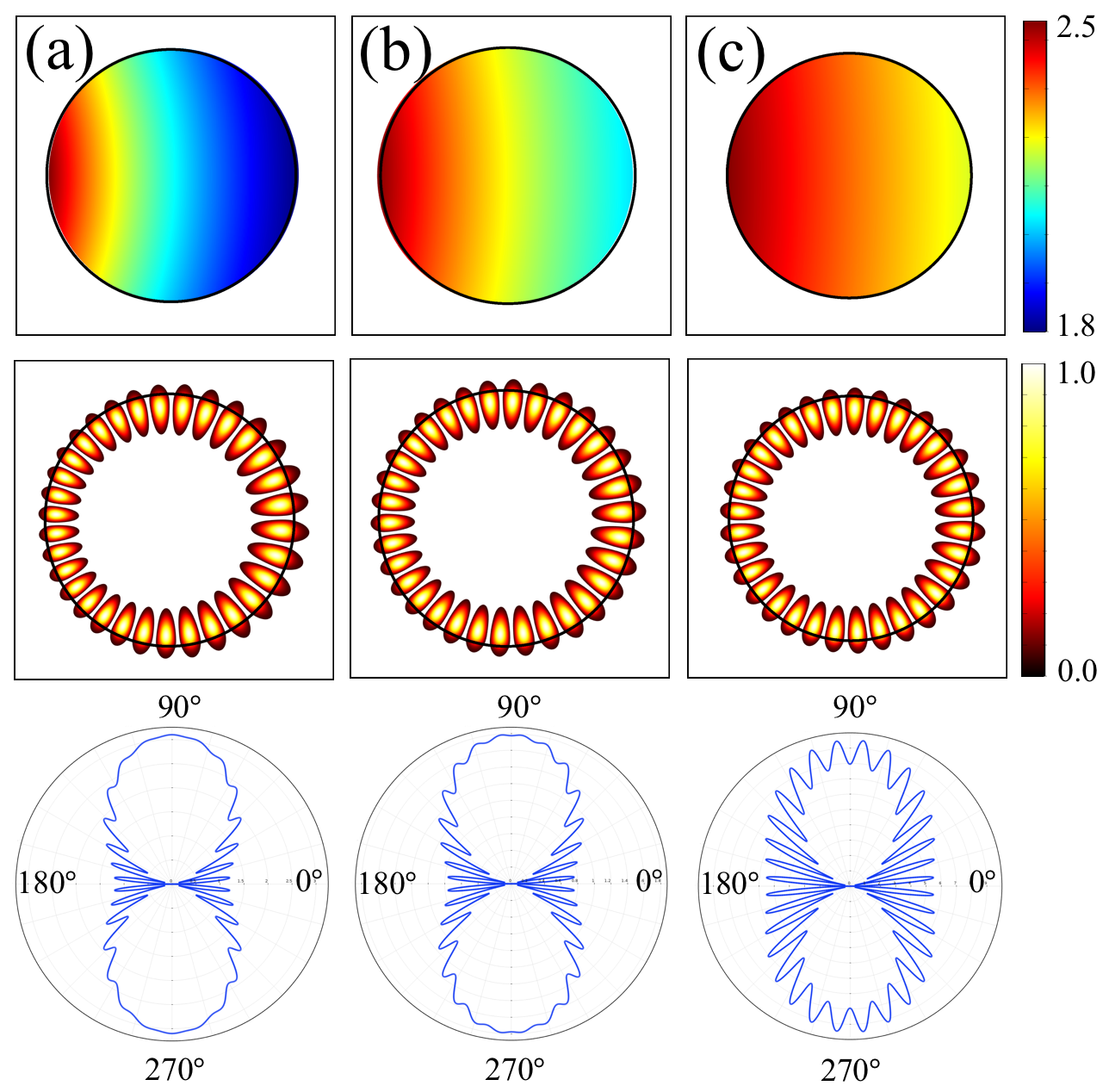}}
\caption{Refractive indices (top), near field intensity patterns (middle), and far field intensity patterns (bottom) of the resonant modes in lima\c{c}on-shaped transformation cavities for (a) ($r$, $\epsilon$) = ($1.0$, $0.081$), (b) ($r$, $\epsilon$) = ($0.4$, $0.125$), and (c) ($r$, $\epsilon$) = ($0.2$, $0.152$).}
\label{fig3}
\end{figure}

We obtained the resonant modes in a TC with a narrow refractive index profile using COMSOL Multiphysics. We start from the WGM with mode numbers $(m,l)=(16,1)$ where $m$ is the azimuthal mode number and $l$ the radial mode number, of which Q-factor is $6967$ in a circular cavity with refractive indices $n_0 = 1.8$ and $n_{out} = 1.0$, respectively, when $r=1.0$ and $\epsilon = 0.0$. The WGM has an isotropic emission because of the rotational symmetry of the circular cavity. As we increase $\epsilon$, the WGM in the circular cavity changes into a cWGM in a lima\c{c}on-shaped TC. Q-factors increase until $\epsilon \sim 0.04$ because of the effect from the higher refractive index and then decrease on account of the breaking of the WGM isotropic properties as $\epsilon$ increases [see black dots in Fig.~\ref{fig2} (a)]. The up-and-down trend of Q-factor is a characteristic of the original design of TCs \cite{Kim16}. Otherwise, the isotropic emission changes into bidirectional emission as $\epsilon$ increases due to the rotational symmetry breaking. As $r$ decreases, however, the decreasing trend of Q-factor as a function of $\epsilon$ disappears, after which the Q-factor increases monotonically as $\epsilon$ increases if $r$ is sufficiently small, as shown in Fig.~\ref{fig2}.

If we consider a material with a refractive index of $n = 2.5$, the maximal value of $\epsilon$ is $\epsilon_c \sim 0.081$ when $r=1.0$. If $\epsilon$ is larger than $\epsilon_c$, the maximal value of the refractive index becomes larger than $2.5$. That is, a TC has a refractive index between $1.8$ and $2.5$ when $r = 1.0$ and $\epsilon = \epsilon_c \sim 0.081$, as shown in Fig.~\ref{fig2} (b). The smaller the parameter $r$, the narrower the width of the refractive index profile. In the $r=0.4$ and $r=0.2$ cases, the corresponding $\epsilon_c$ values are $0.125$ and $0.152$ for $n_{max}=2.5$. Figure~\ref{fig3} shows the TC refractive index profiles, near field intensity patterns, and far field intensity patterns for ($r$, $\epsilon$) = ($1.0$, $0.081$), ($r$, $\epsilon$) = ($0.4$, $0.125$), and ($r$, $\epsilon$) = ($0.2$, $0.152$). The three refractive index profiles all differ but $n_{max} = 2.5$ in all cases. While the near field intensity patterns do not change with different ($r$, $\epsilon$), the far field patterns mostly maintain bidirectional emission when $r \gtrsim 0.3$. As $r$ decreases, the far field patterns smoothly change from bidirectional into isotropic emission. Unlike Q-factor, the definition of "good" emission directionality is not so clear, although there are many useful measures for directional emissions \cite{Son10, Ryu11, Shu13}; such definition depends on the requirements of the particular applications. It is noted that the interference patterns at $90 \degree$ and $270 \degree$ become stronger as $r$ decreases, which are properties of isotropic emission, while there are no interference patterns at $90 \degree$ and $270 \degree$ when emission bidirectionality is clear.

\begin{figure}[tb]
\centering
\fbox{\includegraphics[width=\figsizeone\linewidth]{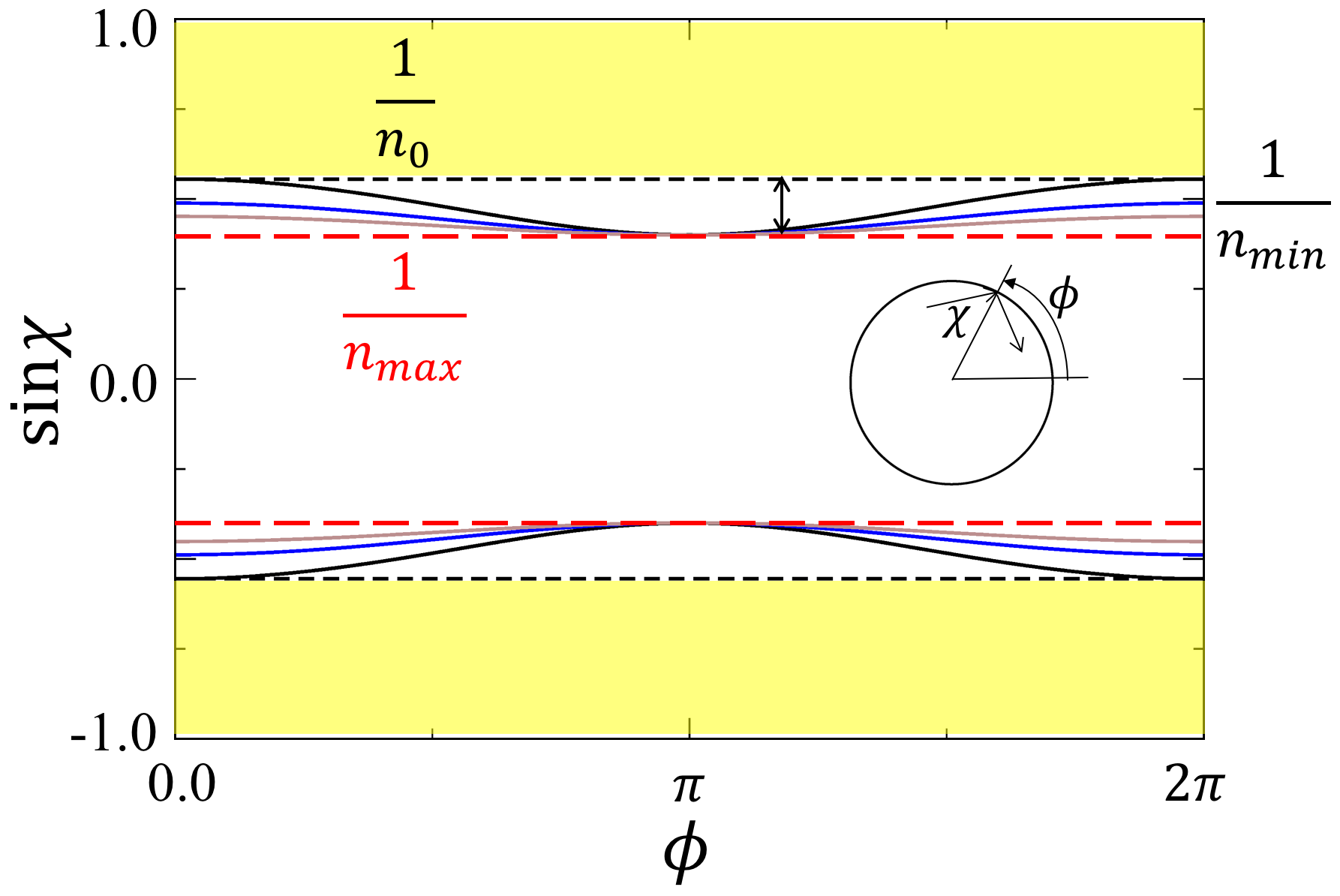}}
\caption{Critical curves for total internal reflection in phase space. The yellow areas and dashed black lines represent the total internal reflection regions and critical lines, respectively, in a circular cavity with a refractive index of $n_{0}$. The black, blue, and brown curves represent critical curves for ($r$, $\epsilon$) = ($1.0$, $0.081$), (b) ($r$, $\epsilon$) = ($0.4$, $0.125$), and (c) ($r$, $\epsilon$) = ($0.2$, $0.152$), respectively. The values of the curves are $1/n_{max}$ at $\phi = \pi$ and $1/n_{min}$ at $\phi = 0$, where $n_{max}$ corresponds to the limitation from the material properties. The black arrow indicates the difference between the reciprocals of the maximal and minimal values of the refractive index profile (see black arrow in Fig.~\ref{fig1}) when $r=1.0$. The inset shows the boundary position $\phi$ and incident angle $\chi$ of the ray trajectory.}
\label{fig4}
\end{figure}

Finally, we discuss why Q-factors increase and emission bidirectionality decreases as $r$ decreases with fixed $n_{max}$ using the phase space of Fig.~\ref{fig4}. In the phase space, $\phi$ and $\chi$ represent the boundary position and the incident angle of a ray trajectory, respectively. Ray trajectories corresponding to WGMs stay in the region of total internal reflection, $\left|\sin \chi \right| > 1/n_0$, and cannot escape from the cavity. The WGMs have a very high Q-factor due to total internal reflection and show isotropic emission because of the rotational symmetry of the circular cavity. While there are critical straight lines for total internal reflection in circular cavities, there are critical curves of which maximal and minimal values are $1/n_{min}$ and $1/n_{max}$ at $\phi = 0$ (or $2 \pi$) and $\phi = \pi$, respectively, when $r=1.0$ and $\epsilon = \epsilon_c = 0.081$. Light emission occurs mainly at the boundary position of $\phi = 0$ (or $2 \pi$) and then generates bidirectional far field patterns at $90 \degree$ and $270 \degree$ because of tangential emission from the boundary. As $r$ decreases with fixed $n_{max}$, when $\sin{\chi} > 0$, the minimal values of the critical curves are fixed as $1/n_{max}$ but the maximal values become smaller, $1/n_{min}$. As a result, the smaller the $r$, the smaller the vertical widths of the critical curves. While the Q-factors of the cWGMs are determined mostly by $1/n_{min}$, emission bidirectionality is related to the difference between $1/n_{min}$ and $1/n_{max}$. As $r$ decreases sufficiently, the critical curves flatten more and the bidirectional emission changes into isotropic emission, like the isotropic emission originating from the relation between the WGMs and critical straight lines in circular cavities.

\section{Summary}
In summary, a new design method has been proposed for a transformation cavity that achieves a gradually varying refractive index profile that is over 50 percent narrower than in the original design. We have studied the Q-factors, near field intensity patterns, and far field intensity patterns of the resonant modes in the transformation cavities. We found that the Q-factor is very high and that the emission directionality survives despite the very narrow refractive index profile in the newly proposed cavities. We expect that such transformation cavities with narrow refractive index profiles will make a breakthrough in the experimental implementation of microcavity lasers employing conformal transformation optics.

\section*{Acknowledgments} The authors thank I. Kim, S. Rim, and M. Choi for the helpful discussions. This research was supported by Project Code (IBS-R024-D1). This research was supported by National Research Foundation of Korea (NRF) grant funded by the Korean government (MSIT) (No. 2020R1A2C3007327 and No. 2020R1A4A1019518).

\end{document}